\documentclass[aps,
prx,
reprint,
showpacs,
twocolumn,
superscriptaddress]{revtex4-2}

\usepackage{graphicx}
\usepackage{dcolumn}
\usepackage{bm}

\usepackage{color}
\usepackage[utf8]{inputenc}
\usepackage[T1]{fontenc}
\usepackage{mathptmx}
\usepackage{etoolbox}
\usepackage{comment}
\usepackage[version=4]{mhchem}
\usepackage{newunicodechar}
\newunicodechar{̈}{\"{}}

\makeatletter
\def\@email#1#2{%
 \endgroup
 \patchcmd{\titleblock@produce}
  {\frontmatter@RRAPformat}
  {\frontmatter@RRAPformat{\produce@RRAP{*#1\href{mailto:#2}{#2}}}\frontmatter@RRAPformat}
  {}{}
}%
\makeatother

\begin{document}

\preprint{AIP/123-QED}

\title{Interplay of Kondo Physics with Incommensurate Charge Density Waves in \ce{CeTe3} }

\author{Aymeric Saunot}
\affiliation{Donostia International Physics Center, 20018 Donostia--San Sebastian, Spain}
\affiliation{Departamento de Polímeros y Materiales Avanzados, Universidad del País Vasco UPV/EHU, 20018 Donostia--San Sebasti\'{a}n, Spain}

\author{Vesna Mik\v{s}i\'{c} Trontl}
\affiliation{Institut za Fiziku, Bijeni\v{c}ka 46, HR--10000 Zagreb, Croatia\\}

\author{Ilya I. Klimovskikh}
\affiliation{Donostia International Physics Center, 20018 Donostia--San Sebastian, Spain}

\author{Denis V. Vyalikh}
\affiliation{Donostia International Physics Center, 20018 Donostia--San Sebastian, Spain}
\affiliation{IKERBASQUE, Basque Foundation for Science, 48013 Bilbao, Spain.}

\author{Alex Louat}
\author{Cephise Cacho}
\affiliation{Diamond Light Source Ltd, Harwell Science and Innovation Campus, Didcot, OX11 0DE, United Kingdom}

\author{Asish K. Kundu}
\affiliation{National Synchrotron Light Source II, Brookhaven National Laboratory, Upton, New York 11973, USA} 

\author{Elio Vescovo}
\affiliation{National Synchrotron Light Source II, Brookhaven National Laboratory, Upton, New York 11973, USA}

\author{Ivana Vobornik}
\affiliation{Istituto Officina dei Materiali (IOM)--CNR, Area Science Park--Basovizza, S.S. 14 Km 163.5, 34149 Trieste, Italy}

\author{Alexander Fedorov}
\affiliation{Helmholtz--Zentrum Berlin für Materialien und Energie, BESSY II, Albert--Einstein--Strasse 15, 12489 Berlin, Germany}
\affiliation{Leibniz--Institut für Festkörper-- und Werkstoffforschung Dresden, Helmholtzstrasse 20, 01069 Dresden, Germany}
\affiliation{Joint Lab Functional Quantum Materials at BESSY II, Albert--Einstein--Strasse 15, 12489 Berlin, Germany}

\author{Cedomir Petrovic}
\affiliation{Condensed Matter Physics and Materials Science Department, Brookhaven National Laboratory, Upton, New York 11973, USA\\}
\affiliation{Shanghai Key Laboratory of Material Frontiers Research in Extreme Environments (MFree), Shanghai Advanced Research in Physical Sciences, Shanghai 201203, China\\}
\affiliation{Vin\v{c}a Institute of Nuclear Sciences, University of Belgrade, Belgrade 11001, Serbia.}

\author{Tonica Valla}
\email{tonica.valla@dipc.org}
\affiliation{Donostia International Physics Center, 20018 Donostia--San Sebastian, Spain}
\affiliation{Institut za Fiziku, Bijeni\v{c}ka 46, HR--10000 Zagreb, Croatia\\}

\date{\today}

\begin{abstract}
\ce{CeTe3} is a 2--dimensional (2D) Van der Waals (VdW) material with incommensurate charge density waves (CDW), extremely high transition temperature ($T_{CDW}$) and a large momentum--dependent CDW gap that leaves a significant portion of the Fermi surface intact. It is also considered to be a weak Kondo system, a property unexpected for a material with incommensurate CDW, where each atomic site is slightly different. Here, we study the properties of the CDW state in several \ce{RTe3} (R is rare earth) materials and examine the hybridization of itinerant states with the localized Ce $4f$ multiplet in \ce{CeTe3} by using angle resolved photoemission spectroscopy (ARPES). We find that the renormalization of the itinerant states originating from the hybridization with the deeper localized $4f$ states at $-260$ meV is $k-$dependent and extends to the Fermi level. As these localized states are far from the Fermi level, the observed hybridization affects the effective masses only marginally and does not lead to heavy fermions. However, since the same renormalizing mechanism normally leads to the heavy fermion physics when the localized $4f$ states are near the Fermi level, our observation of its strong $k-$dependence suggests that this could be the reason for discrepancy between the heavy masses in specific heat and light ones in Shubnikov de Haas oscillations, often observed in heavy fermions. 
\end{abstract}

\maketitle


In \ce{RTe3}, where R is a rare earth, CDW is incommensurate and very strong, with extremely high transition temperatures ranging up to $>600$ K and large gaps, of the order of few hundred meV.\cite{DiMasi1995,Brouet2008,Tomic2009,Hu2014,Ralevic2016,Yumigeta2021,Sarkar2023,Regmi2023} The exact mechanism of the CDW formation is not entirely known, but the Fermi surface (FS) nesting is thought to play a much more significant role than in transition metal dichalcogenides, where the strong anisotropic electron--phonon coupling is necessary to drive the system in the CDW state. \cite{Varma1983,Valla2004,Kundu2024} Although some studies suggest that the strong electron--phonon coupling is the dominant player, \cite{Eiter2013,Maschek2015,Hong2022} other studies indicate that the coupling is relatively weak in \ce{RTe3} materials.\cite{Brouet2008,Liu2020,STM2024} Irrespective of the mechanism, once the CDW is established, the perfect 4--fold symmetry of the normal state ($a=c$) is broken as the CDW runs either uniaxially along the $c$ crystallographic axis for La--Tb, or biaxially, but with two unequal wave vectors and amplitudes, for the Dy--Tm compounds. \cite{Yumigeta2021} This results in slight orthorhombicity, where $a$ and $c$ lattice constants differ by  ~0.2--0.3 \% in the CDW state. \cite{Malliakas2005} Both the uniaxial and biaxial CDWs are incommensurate with the lattice, implying that every lattice site is slightly different in terms of local charge density. \cite{DiMasi1995,Yumigeta2021} 

In addition to CDW, \ce{CeTe3} is also thought to display a weak Kondo behavior. \cite{Ru2006,Brouet2008} Normally, the CDW and Kondo physics would be mutually exclusive. Not only that CDW reduces the density of states at the Fermi level necessary for screening of local moments, but the Kondo physics requires a screening cloud that should screen all the local moments, producing the heavy fermion (HF) ground state with local singlets. If the CDW is truly incommensurate, every localized moment feels slightly different environment and its screening might require somewhat different Kondo cloud. The Kondo clouds have never been seen experimentally in real materials, but the theoretical considerations indicate that they have to be large, on the micron scale. \cite{Affleck2010,Park2013a,Shim2023} This has recently been verified in the arrays of quantum dots. \cite{V.Borzenets2020} In materials with incommensurate CDW, such as \ce{CeTe3}, the enormous size of the cloud might be detrimental to Kondo screening and the HF state should not form. If it does, \ce{CeTe3} would be only the second VdW HF system, in addition to recently discovered \ce{CeSiI}. \cite{Jang2022,Posey2024} 

Another fundamental problem is that some HF materials show heavy carrier masses in specific heat and very light carriers in quantum oscillations in transport. \cite{Duality115,Mun2013,Mun2015,Posey2024}
Two different explanations for this peculiar duality were discussed. First, it was suggested that the high magnetic field required for quantum oscillations suppresses the Kondo screening, turning the carriers light. \cite{Duality115,Mun2013,Posey2024} 
The second proposal is that the hybridization of the itinerant carriers with the localized Ce $4f$ moments might be $k-$dependent, leading to simultaneous existence of heavy and light bands. \cite{Vyalikh2010,Posey2024} However, studies to adequately quantify these effects, are still lacking. 

Here we perform ARPES studies on \ce{CeTe3} and on iso--electronic \ce{LaTe3} and \ce{GdTe3} compounds and characterize the features of their electronic structure related to the incommensurate CDW. In \ce{CeTe3}, we detect traces of hybridization of the localized Ce 4$f-$derived state at the Fermi level with the highly itinerant states that remain ungapped and form the FS in the CDW state. We also find that the hybridization of the itinerant states with the deeper multiplet of Ce $4f$ localized states is strongly $k-$dependent. This indicates that: 1) the Kondo physics can play a role in materials with incommensurate CDW and 2) the $k-$dependent hybridization of itinerant and localized states may be responsible for the dualities seen in HFs, where different experimental techniques see different masses of carriers. The latter is unrelated to CDW, and should be generally consider for the HF systems. These findings answer some of the most fundamental questions related to HF materials.



\begin{figure*}[htpb]
\begin{center}
\includegraphics[width=15cm]{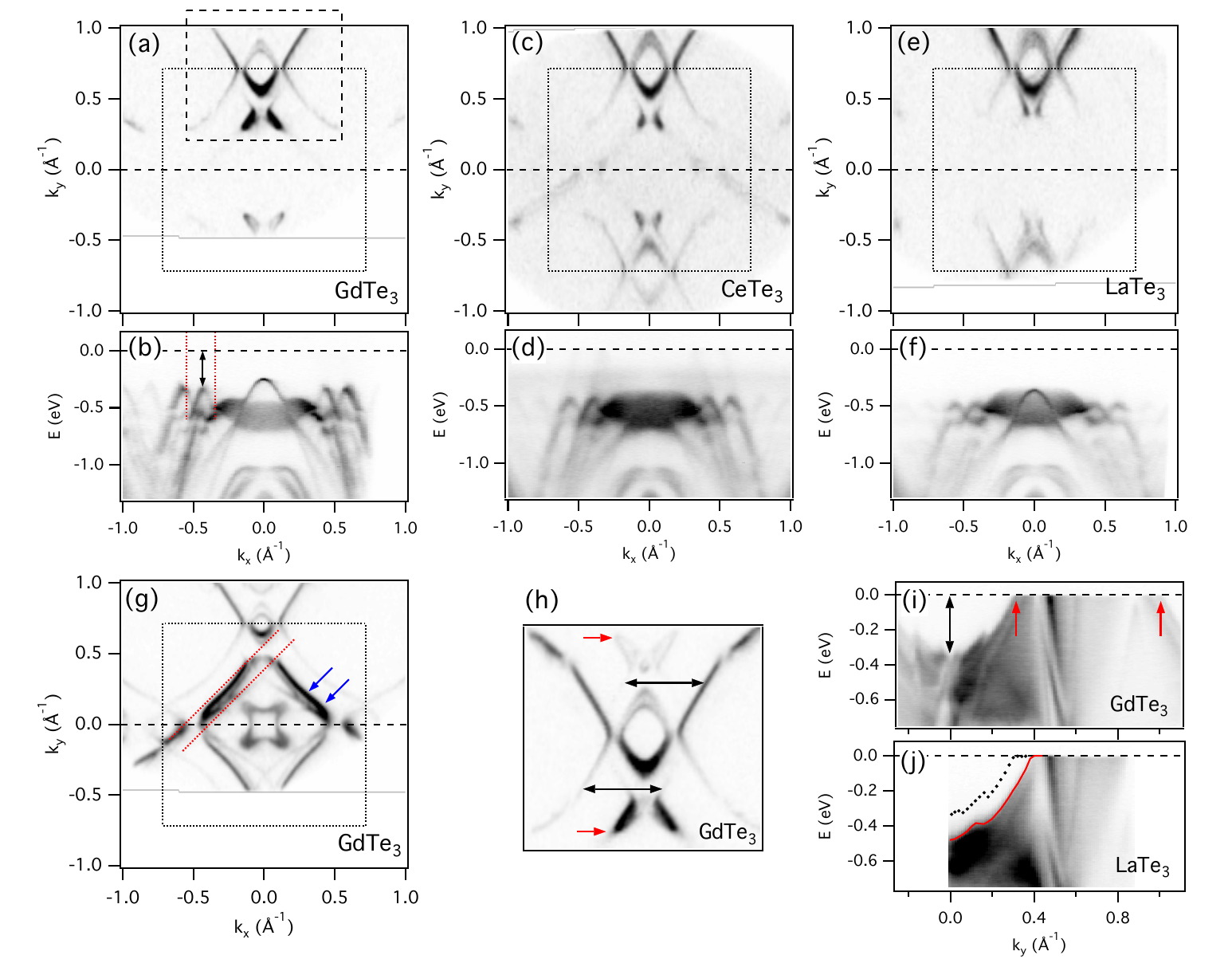}
\caption{Charge density waves in three \ce{RTe3} materials. (a) The spectral intensity at $E=0\pm4$ meV, the Fermi surface (FS) and (b) bands dispersions along the $k_y=0$ line in the Brillouin zone (BZ) for \ce{GdTe3}. (c,d) The same for \ce{CeTe3} and (e,f) for  \ce{LaTe3}. (g) Spectral intensity at $-0.35$ eV for \ce{GdTe3}. The dotted black squares in (a,c,e,g) represent the first BZ. (h) Zoomed in region of the \ce{GdTe3} FS marked by the dashed rectangle in (a). (i) Spectral intensity integrated over the $k$ region marked by red dotted lines in (b,g) in \ce{GdTe3} and (j) in \ce{LaTe3}. The black circles and the red curve in (j) represent the leading edge positions of the integrated intensity for \ce{GdTe3} and \ce{LaTe3}, respectively. The black arrows in (b,i) represent the maximal CDW gap on the occupied side of the spectrum at $k_y=0$. The dotted red lines in (b,g) represent the integration region used for the extraction of $\Delta_{CDW}(k_y)$ shown in (i,j). Blue arrows in (g) indicate the points in the BZ where $\Delta_{CDW}$ protrudes into the otherwise monotonic $k_y$ dependence shown in (i). The black arrows in (h) represent the $q_{CDW}$. The red arrows in (h,i) indicate the $k_y^{crit}$ dividing the gapped and ungapped regions of the BZ. All the spectra were acquired at $h\nu=70$ eV and 10 K.
}
\label{Fig1}
\end{center}
\end{figure*}

Fig. \ref{Fig1} shows the FSs and some important details of the electronic structure of \ce{GdTe3}, \ce{CeTe3} and \ce{LaTe3} in the CDW state. Although the direct comparison indicates remarkable similarities between these compounds, it also uncovers some important differences. Firstly, it is obvious that the FS is uniaxially gapped by CDW in all three compounds. The CDW completely gaps the region within $|k_y|\leq k_a^{crit}$, while the rest of the FS is kept intact, breaking the tetragonal symmetry of the normal state. Some of the fundamental parameters of CDW state can be directly extracted from the ARPES data: for example the CDW gap magnitude on the occupied side is found to be increasing significantly from \ce{GdTe3} to \ce{LaTe3} as indicated in Fig. \ref{Fig1}(b,d,f,i,j). Further, the CDW wave vector, $q_{CDW}$, can also be extracted directly from ARPES measurements as indicated in Fig. \ref{Fig1}(h). Although our values for $q_{CDW}$ are not as precise as those obtained by x--ray diffraction, they are in general agreement with them, indicating the incommensurate CDW character with $q_{CDW}\approx2/7(2\pi/c)$. \cite{DiMasi1995}
Lastly, the boundaries of the Brillouin zone affected by the CDW, $k_a^{crit}$, can also be precisely measured in ARPES, as shown in Fig.\ref{Fig1}(h--j). Table \ref{tab} summarizes the parameters extracted directly from our ARPES data. 

The magnitudes of the occupied part of the gap agree well with the recent ARPES studies. \cite{Sarkar2023,Regmi2023} When comparing those values with the STM and optical data, one should take into account that the CDW gaps are not particle-hole symmetric and that ARPES only provides the portion of the gap on the occupied side. Also, STM and optical probes are generally not $k-$resolved, and due to the fact that the gap is strongly $k-$dependent and does not affect some states at all, makes the comparison even more complicated.  However, the trend across different compounds is universal. \cite{Brouet2008,Tomic2009,Hu2014,Ralevic2016}

In \ce{LaTe3} the gap is the largest, $\Delta_{CDW}^{occ}=0.48$ eV, and it affects the largest portion of the BZ, both contributing to one of the most robust CDW states in condensed matter, with extremely high CDW transition temperature, $\sim$600 K.

The detailed $k-$dependence of the gap along the state that would have otherwise formed the FS is shown in Fig.\ref{Fig1}(i,j). It is obviously unidirectional in all three compounds, depending only on $k_y$.  We note that the recent $k_x, k_y$ (2D) dependence introduced by Regmi \textit{et al} \cite{Regmi2023} can easily be converted to the unidirectional one, depending only on $k_y$. However, the $k_y$ dependence is not completely monotonic as suggested by previous studies, \cite{Brouet2008,Sarkar2023,Regmi2023} but shows some fine structure, with the spikes of intensity protruding into the otherwise smooth gap curve at several $k_y$ points, as best seen in \ce{GdTe3} (Fig. \ref{Fig1}(i)). These spikes originate from avoided crossings of the gapped bands at these momenta, as indicated in Fig. \ref{Fig1}(g). 

It is interesting that within the gapped region of the BZ, there are certain states that go above the  $\Delta_{CDW}^{occ}$ at a given $k_y$. For example, the sharp parabolic state at the $\Gamma$ point is always above the maximal $\Delta_{CDW}^{occ}$, most notably in \ce{GdTe3} and \ce{LaTe3} (Fig.\ref{Fig1}(b,f)). This is because these states cannot be effectively nested by $q_{CDW}$ and therefore remain unaffected, similarly to the unaffected FS at $|k_y|> k_a^{crit}$.

Finally, we briefly comment on the importance of the electron--phonon coupling in driving the \ce{RTe3} into CDW state. The spectral width of electronic states forming the FS, both as a function of energy and temperature, suggests that the coupling is weak in all three compounds. \cite{Valla1999a} It is much weaker than in transition--metal dichalcogenides, such as 2H--NbSe$_2$ and 2H--TaSe$_2$, materials with less robust CDW, but with the strong and anisotropic coupling thought to play a significant role in CDW. \cite{Valla2000b,Valla2004,Arguello2015,Kundu2024}In fact, the coupling is so weak that it is difficult to quantify, with no apparent kinks in the state's dispersions that could point to the structure in electronic self--energy, $\Sigma(\omega)$, similar to topological insulators. \cite{Pan2012,Kundu2025} The absence of kinks could still allow for strong coupling to low--frequency phonons, beyond our resolution, $\omega_0<6$ meV. Such coupling would then cause broadening of the FS $\sim\lambda\pi k_B T$ nearly linear at temperatures above $\sim k_BT>\omega_0/3$, where $k_B$ is the Boltzmann’s constant and $\lambda$ is the electron--phonon coupling constant.\cite{Valla1999a} 
However, our temperature dependence shows no apparent broadening of the states as the temperature is raised to $\sim150$ K, in accord with the low Debye and superconducting transition temperatures under pressure, \cite{Straquadine2022,Ru2006,Zocco2015} and with the recent STM results that estimate the coupling to be $\lambda\approx0.19$. \cite{STM2024} This suggests that electron--phonon coupling plays a relatively minor role in CDW and accentuates the importance of FS nesting, similar to the situation in another group of square--net materials based on Sb. \cite{Bosak2021,Wu2023}

\begin{table}
\caption{\label{tab}CDW parameters for \ce{GdTe3}, \ce{CeTe3} and \ce{LaTe3} from Fig.\ref{Fig1}}
\begin{ruledtabular}
\begin{tabular}{lccccc}
Sample & $\Delta_{CDW}^{occ}$ (eV)& $q_{CDW}$ ($2\pi/c$)& $k_a^{crit}$ ($2\pi/a$) \\
\colrule
\ce{GdTe3} & 0.333 & 0.285 & 0.205\\
\ce{CeTe3} & 0.441 & 0.276 & 0.227\\
\ce{LaTe3} & 0.480 & 0.275 & 0.267\\
\end{tabular}
\end{ruledtabular}
\end{table}


\begin{figure*}[htpb]
\begin{center}
\includegraphics[width=14
cm]{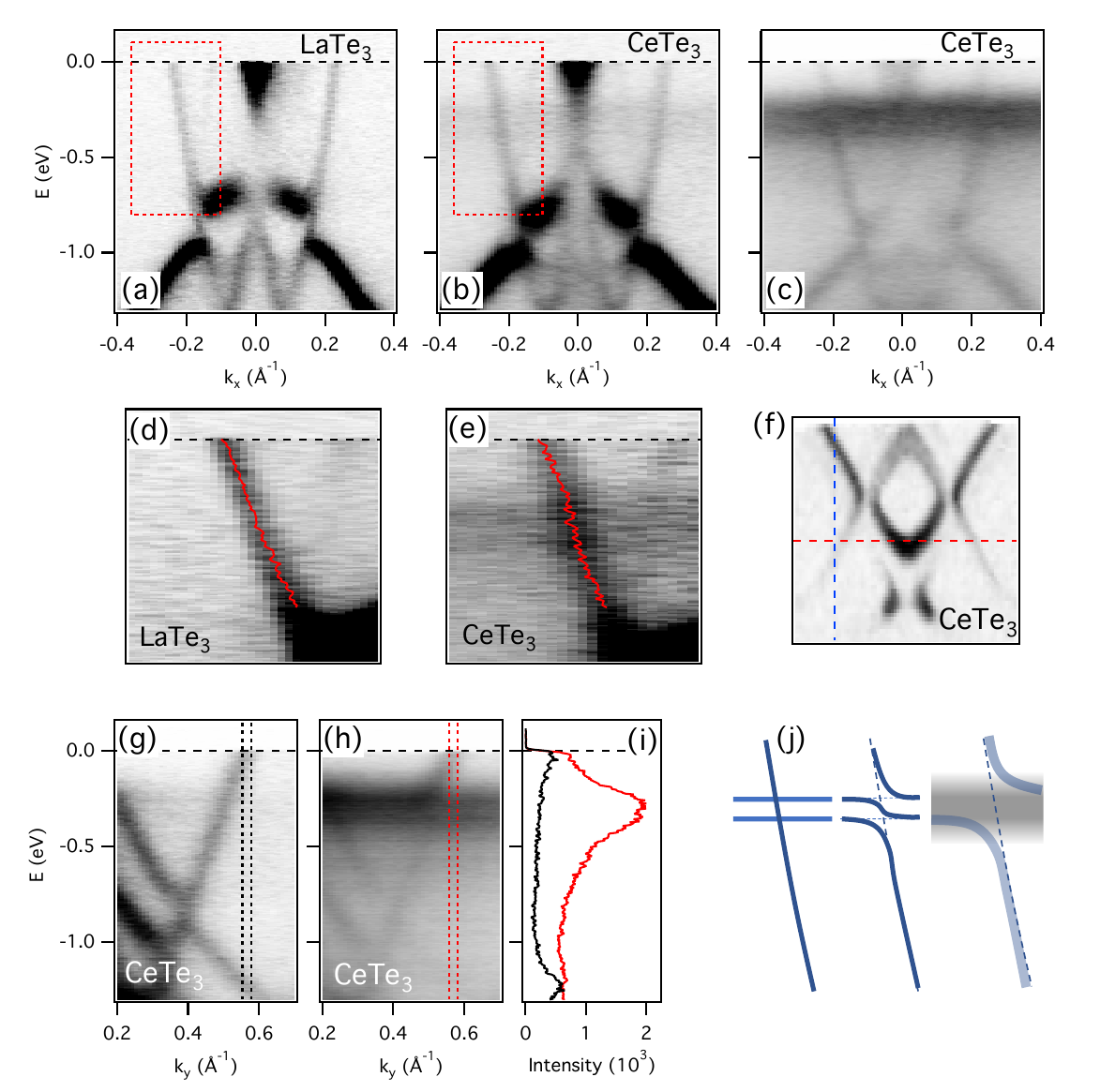}
\caption{Hybridization of itinerant states with localized $4f$ states in \ce{CeTe3}. (a) Photoemission intensity along the red dashed line in (f) for \ce{LaTe3} taken at $h\nu=70$ eV. (b) The same for \ce{CeTe3}. (c) The same, but recorded at $h\nu=121$ eV. (d) and (e) close--ups of spectral intensity from the red dashed boxes in (a) and (b), respectively. Red curves represent dispersions obtained by fitting the momentum distribution curves (MDC) by a Lorentzian peak on a linear background. (f) The FS of \ce{CeTe3} with the momentum lines probed in (a--e) (red) and (g--i) (blue). (g) Spectral intensity for \ce{CeTe3} along the blue dashed line in (f) taken at $h\nu=115$ eV. (h) The same , but recorded at $h\nu=121$ eV. (i) EDCs around $k_F$ for the two spectra shown in (g, h). (j) Schematic of the interaction of an itinerant band with the localized state(s). No hybridization (left), hybridization with a doublet turned on (center) and hybridization with a broad localized continuum (right). 
}
\label{Fig2}
\end{center}
\end{figure*}

It is already visible in Fig. \ref{Fig1} that while Gd-- and La-- based materials show a very clean gap at $k_y=0$, \ce{CeTe3} shows a non--dispersing intensity inside the gap, peaked around $-0.26$ eV. This intensity originates from the localized Ce $4f-$derived multiplet of spin--orbit shake--off states, as further evidenced in Fig. \ref{Fig2}. Fig. \ref{Fig2} shows several details of the electronic structure of \ce{CeTe3} when the enhancement of photoemission near the $4d-4f$ resonance is turned \textit{on} and \textit{off}. The itinerant bands that form the FS are nearly identical in \ce{LaTe3} and \ce{CeTe3} and they remain mostly unchanged when the resonant photoemission is turned \textit{on} in \ce{CeTe3}, but the intensity of localized Ce $4f-$derived states is strongly enhanced. This is visible in the 2D images of the band dispersions, Fig. \ref{Fig2}(c) and (h), and in the energy distribution curves, Fig. \ref{Fig2}(i). The most of the enhancement is around $\sim 260$ meV below the Fermi level. We assign this to the $4f$ spin--orbit shake--off multiplet, or the $4f^1_{7/2}$ state in the traditional nomenclature. \cite{Allen1986,Patil2016,CeCoIn,Posey2024} Our spectra suggest that the multiplet is at least a doublet, and that due to its significant width, it contributes to some intensity at the Fermi level. However, as the Fig. \ref{Fig2}(i) illustrates, there might be an additional enhancement of the photoemission signal near the Fermi level that does not originate from the spectral structure at  $-260$ meV. In Ce systems, the emission near the Fermi energy primarily arises from the $4f^1_{5/2}$ peak that overlaps with the Kondo resonance. \cite{Allen1986,Patil2016,CeCoIn,Poelchen2020,Posey2024} Its weak intensity indicates that the Ce 4$f$ moments interact relatively weakly with the itinerant states. This suggests that \ce{CeTe3} is a weak Kondo system, in agreement with the transport measurements reported by Ru \textit{et al}. \cite{Ru2006}

\begin{figure*}[htpb]
\begin{center}
\includegraphics[width=15
cm]{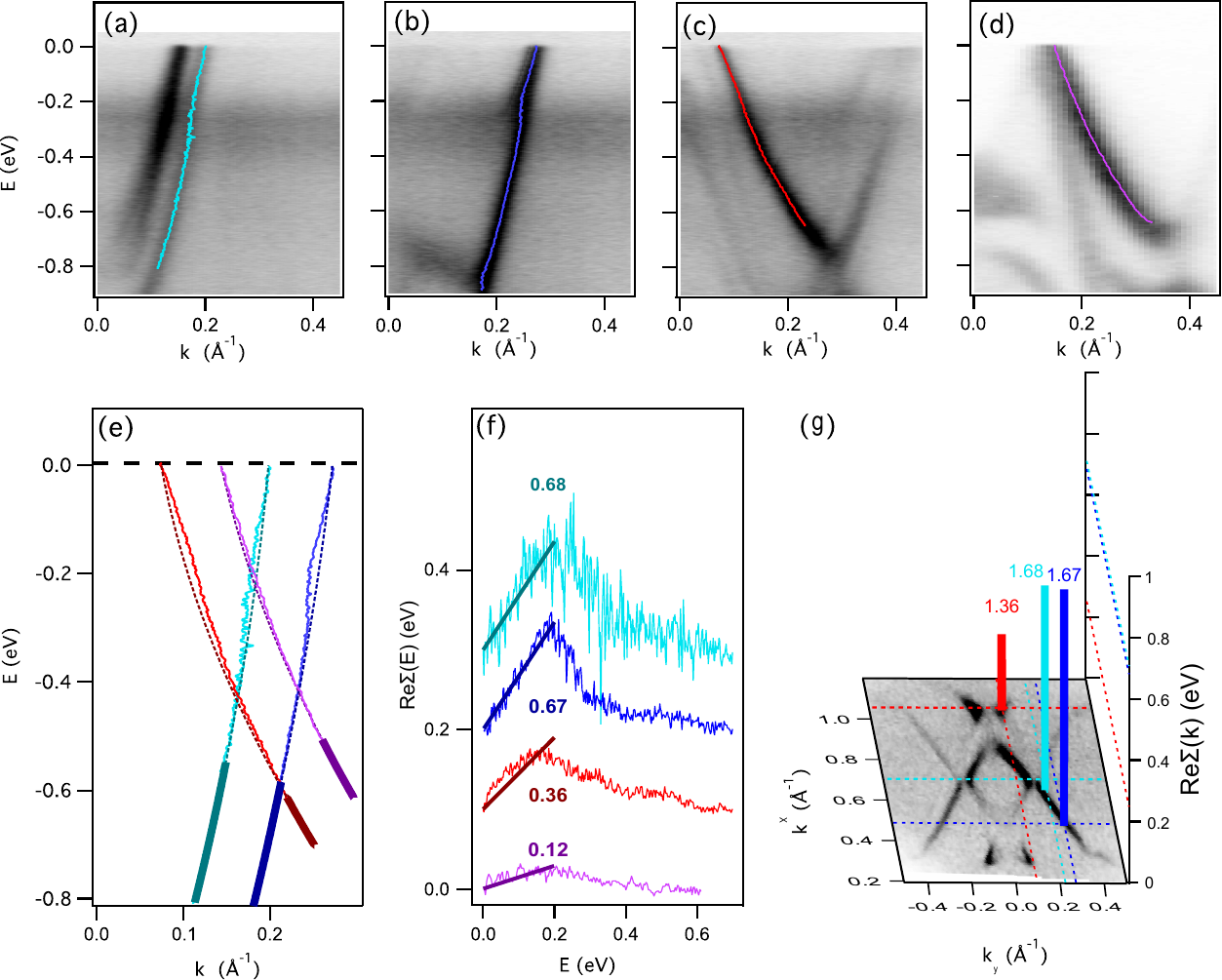}
\caption{$k-$dependent hybridization of itinerant and localized states in \ce{CeTe3}. (a-c) Photoemission intensity along several momentum lines for \ce{CeTe3} as indicated in (h). (d) The photoemission intensity for \ce{GdTe3} along the red dashed line in (g). The dispersions derived from fitting of MDCs by Lorentzian peaks are represented by different colors. (e) The same measured dispersions from (a-d), with the "non-interacting" bands (dotted curves), obtained by fitting the measured dispersions with the 2$^{nd}$ degree polynomial in the high-energy range as indicated by the thick solid curves). 
(f) $Re\Sigma(E,k)$ for the bands from (e). The linear fit to the low energy part represents the coupling strength $\lambda$, that determines the mass enhancement factor $m^*/m=1+\lambda$.
(g) FS of \ce{CeTe3} with the momentum lines probed in (a--c) indicated by the dashed lines. Also plotted are the mass enhancement factors $m^*/m=1+\lambda$.
}
\label{Fig3}
\end{center}
\end{figure*}

Also evident from Fig. \ref{Fig2} is that the itinerant states that cross the multiplet at 260 meV, hybridize with it and become "heavier", i.e. their velocity is re--normalized. This is the main difference compared to \ce{GdTe3} and \ce{LaTe3}: while the state in Fig. \ref{Fig2}(a,d) is essentially a straight line, the corresponding state in \ce{CeTe3} has a kink in dispersion at the energy where it crosses the $4f$ multiplet (Fig. \ref{Fig2}(b, c, e, g, h). This is particularly obvious at the resonant photoemission, Fig. \ref{Fig2}(c) and (h). The observation can be qualitatively described by the cartoon in Fig. \ref{Fig2}(j): the itinerant states hybridize with a localized multiplet, or with a broad localized continuum and get renormalized over the energy range over which the multiplet (or a continuum) is spread. Here, this means that even at the Fermi level there will be a finite mass renormalization -- the carriers at the FS will be heavier in \ce{CeTe3} than in La-- or Gd--based compounds. 

We now turn to the $k-$dependence of the observed hybridization and the resulting $k-$dependent renormalization of the Fermi velocity in \ce{CeTe3}. As mentioned before, such a $k-$dependence could possibly explain the dichotomy between the carrier's masses probed in transport and specific heat, often seen in HF materials.
In Fig. \ref{Fig3}, we show dispersions along several different momentum lines in \ce{CeTe3} as indicated in panel (g). The spectra in panels (a-c) were performed at $\sim2-3$ K, in the regime where the Kondo effect is both present and coherent. \cite{zeng_kondo-coupled_2025} We note that there are no observable changes in the spectra upon lowering temperature from $\sim 10$ to $\sim2-3$ K. However, it is obvious that the hybridization of the itinerant states with the localized $-260$ meV multiplet is not the same in all the spectra. The dispersions extracted by Lorentzian fitting of momentum distribution curves (MDC) are also shown and are replotted in Fig. \ref{Fig3}(e).\cite{Valla1999a} 

Next, in order to quantitatively describe the hybridization of the itinerant bands with the localized \ce{Ce} $4f$ multiplet, we plot the analogue of the real part of the self-energy, Re$\Sigma(E,k)$. We emphasize that the extracted quantity is not really Re$\Sigma$, as it would be in the case of electron-phonon interaction, where it would be causally related to Im$\Sigma$.\cite{Valla1999a} Here, it represents a quantity that quantitatively describes mass enhancement of an itinerant band, resulting from hybridization on the Ce $4f$ multiplet. To extract Re$\Sigma$, we fitted the measured band dispersion at higher energies, far from the Ce $4f$ state, using a second-order polynomial forced to go through $k_F$ (Fig. \ref{Fig3}(e)). We used this polynomial as a "non-interacting" band dispersion. Then, the difference in momentum, $\Delta k(E)$, between the measured dispersion and the "non-interacting" one is multiplied by the group velocity of the "non-interacting" band, resulting in Re$\Sigma(E,k)=-v(E,k)\Delta k(E,k)$ (Fig. \ref{Fig3}(f)). The extracted Re$\Sigma$ has a pronounced peak at $\sim 200$ meV, reflecting the renormalization of mass and velocity at the localized $4f$ multiplet. 
Its low-energy slope was then extracted by fitting Re$\Sigma$ to a straight line, to obtain the mass-enhancement parameter, $\lambda$.
We note that, if the same procedure is performed for itinerant bands in Gd$Te_3$ (Fig. \ref{Fig3}(d))and La$Te_3$, the resulting Re$\Sigma$ and $\lambda$ both tend to vanish, reflecting the lack of $4f$ states and assuring that our analysis renders reasonable results.

The crucial new result here is that, although the mass enhancement is generally small, it varies around the FS: the renormalization is $k-$dependent! Therefore, we have demonstrated that hybridization between the itinerant states and the localized $4f$ moments, the essential part of the Kondo physics that, in its lattice version, results in HF character, can indeed be $k-$dependent. 
Although this possibility has been suggested before, it has never been quantitatively demonstrated. \cite{Vyalikh2010,Posey2024}

Recent optics studies have reported significantly larger mass enhancement factors, of the order of $\sim10$.\cite{zeng_kondo-coupled_2025} Optics averages over the FS, and the fact that it still returns a factor 5-6 times larger than our strongest $k-$resolved renormalization indicates that there might be additional renormalization at a $4f$ resonance much closer to the Fermi level, as Fig. \ref{Fig2}(i) might imply. Although our ARPES studies do not have enough resolving power to detect such hybridization, we anticipate that the renormalization on this state could have been much stronger, and that the carrier mass would vary more drastically around the FS.

We argue that the observed $k-$dependence would have profound effect on transport properties: the weakly hybridized states (Fig. \ref{Fig3}(c)) would likely dominate the charge transport and contribute to "light" carrier character, whereas the strongly hybridized ones (Fig. \ref{Fig3}(a,b)) would dominate specific heat and manifest the "heavy" carrier character. That would explain the standing controversy in some HF materials. \cite{Duality115,Vyalikh2010,Posey2024} The other effect probably contributing on the same footing is a suppression of the Kondo screening in strong magnetic field that is needed for transport quantum oscillations. \cite{Mun2013}


We have performed ARPES studies on \ce{CeTe3}, \ce{LaTe3} and \ce{GdTe3} and characterized various aspects of their electronic structure related to the incommensurate CDW. In \ce{CeTe3}, we have detected hybridization of the highly itinerant states forming the FS in the CDW state with the localized Ce 4$f-$derived states. The localized states form a dominant broad multiplet centered around 260 meV below the Fermi level with  a possible weak feature very near the Fermi level. We have found the strong evidence of the $k-$dependent hybridization of the itinerant states with the deeper multiplet of Ce $4f$ localized states. 
Our results imply that the basic Kondo physics can certainly survive in materials with incommensurate CDW. However, the fact that \ce{CeTe3} displays a weak Kondo behavior, rather than the HF one, might be a sign that the screening is indeed affected. Equally important is our finding that the hybridization of itinerant and localized states is $k-$dependent. The same phenomenon may be responsible for apparent discrepancies observed in some actual HF materials, where different probes see different effective masses of carriers.


Single crystals were grown via a self--flux technique as described by Ru \textit{et al}. \cite{Ru2006}
The ARPES experiments were carried out at four different beamlines (the I--05 beamline at Diamond, the ESM beamline at NSLS II, UE112--PGM2b $1^3$ beamline at BESSY and the APE--LE beamline at Elettra) either with the MBS or Scienta DA--30 electron spectrometers and photons in the range from 50 to 140 eV. The total instrumental energy resolution was $\sim$ 5--6 meV at I--05, ESM and UE112--PGM2a, and  $\sim$ $12-18$ meV at APE--LE. The angular resolution at all facilities was better than $\sim 0.15^{\circ}$ and $0.3^{\circ}$ along and perpendicular to the entrance slit of the analyzer, respectively.\\

Data availability

The data that support the findings of this study are available from the corresponding author upon reasonable request. 

Acknowledgments

T.V. and I.I.K. acknowledge the support from the Red guipuzcoana de Ciencia, Tecnología e Innovaci\'{o}n – Gipuzkoa NEXT 2023 from the Gipuzkoa Provincial Council under Contract No. 2023--CIEN--000046--01. V.M.T. and T.V. were supported by the Centre for Advanced Laser Techniques (CALT), grant KK.01.1.1.05.0001. I.I.K was supported by the fellowship from la Caixa Foundation (ID  100010434) under the code LCF/BQ/PI24/12040021.
CP was supported by Shanghai Key Laboratory of Material Frontiers Research in Extreme Environments (MFree), China (no. 22dz2260800), Shanghai Science and Technology Committee, China (no. 22JC1410300). 
The ARPES work was carried out with the support of Diamond Light Source, instrument i05 (proposal SI36637--1), the National Synchrotron Light Source II, the ESM beamline (proposal 315476), Elettra, APE--LE beamline (proposal 20235130) and BESSY, UE112--PGM2a beamline (proposal 232--12385--ST).
Work at Brookhaven National Laboratory was supported by U.S. DOE, Office of Science, Office of Basic Energy Sciences under Contract No. DE--SC0012704. 

Author information

Author's Contributions:
T.V. directed the study, analyzed and interpreted data and wrote the manuscript. A.S. analyzed the data. V.M.T., I.I.K., A.K.K., D.V.V., A.S. and T.V. performed the ARPES experiments. A.L., C.C., E.V. I.V. and A.F. supported the ARPES experiments at synchrotron facilities. C.P. grew the crystals. All the authors commented on the manuscript.

Competing interests:
The authors declare no competing interests.

\section*{References}


%

\end{document}